# DataCite as a novel bibliometric source: Coverage, strengths and limitations

Nicolas Robinson-Garcia[1]*, Philippe Mongeon[2], Wei Jeng[3] and Rodrigo Costas[4,5]

[1]INGENIO (CSIC-UPV), Universitat Politècnica de València (Spain)
elrobin@ingenio.upv.es
* Corresponding author
[2]École de bibliothéconomie et des sciences de l'information, Université de Montréal (Canada)
philippe.mongeon@umontreal.ca
[3]Department of Library and Information Science, National Taiwan University, (Taiwan)
wjeng@ntu.edu.tw
[4]CWTS, Leiden University (The Netherlands)
rcostas@cwts.leidenuniv.nl
[5]Centre for Research on Evaluation, Science and Technology (CREST), Stellenbosch University, Private Bag X1, Matieland 7602 (South Africa)

## Abstract
This paper explores the characteristics of DataCite to determine its possibilities and potential as a new bibliometric data source to analyze the scholarly production of open data. Open science and the increasing data sharing requirements from governments, funding bodies, institutions and scientific journals has led to a pressing demand for the development of data metrics. As a very first step towards reliable data metrics, we need to better comprehend the limitations and caveats of the information provided by sources of open data. In this paper, we critically examine records downloaded from the DataCite's OAI API and elaborate a series of recommendations regarding the use of this source for bibliometric analyses of open data. We highlight issues related to metadata incompleteness, lack of standardization, and ambiguous definitions of several fields. Despite these limitations, we emphasize DataCite's value and potential to become one of the main sources for data metrics development.

## 1. Introduction

Calls for data availability and sharing can be traced back to the beginning of the 20th century when Galton stated: "I have begun to think that no one ought to publish biometric results, without lodging a well arranged and well bound manuscript copy of all his data, in some place where it should be accessible, under reasonable restrictions, to those who desire to verify his work" (Galton, 1901, as cited in Perneger, 2011). However, it has been just a few decades since technology has made possible the development of the necessary infrastructure to make this happen (Peng, 2011). In the last decade, public funding agencies, publishers and institutions have directed their efforts towards developing such infrastructure as well as to incentivizing data sharing and reuse within the scientific community by promoting data citations (Robinson-García, Jiménez-Contreras, & Torres-Salinas, 2015).

Data sharing and reuse practices have been adopted at a different pace by the different scientific communities. For instance, data infrastructure is widely developed within the crystallography community, dating back to the early 1970s (Torres-Salinas, Robinson-García, & Cabezas-Clavijo, 2012). A similar expansion can be observed in Genomics or Astronomy (Borgman, 2012). On the other hand, social sciences and the humanities have thus adopted these new practices at a slower pace than STEM fields (Doorn, Dillo, & van Horik, 2013; Kim & Adler, 2015).

Infrastructure design is a key factor towards fostering data sharing and reuse. Piwowar, Becich, Bilofsky and Crowley (2008) analyzed how certain elements of data sharing frameworks may influence the usability, discoverability, and data reuse for different stakeholders.

Although measuring the impact of data is a highly relevant element in the research policy agenda, a direct measure of data reuse is very difficult to achieve (Missier, 2016). Attempts of metrics such as downloads of datasets or data citations have been proposed to track data reuse (Konkiel, 2013). While the former seem to be problematic on capturing different dimensions of usage (Mayernik, Hart, Maull, & Weber, 2016), -- e.g., data might be downloaded for research validating purposes, -- more effort has been put into the call of movement of "data citations" (Costas, Meijer, Zahedi, & Wouters, 2013; Piwowar, Day, & Fridsma, 2007).

For data citations to become a valid indicator on data reuse, a shift is needed on the communication behavior of researchers when citing sources, as well as on the meaning they attach to their references (Mayernik, 2012; Parsons & Fox, 2013). Initiatives such as the launch of the Data Citation Index and the DataCite consortium are examples of efforts directed at promoting data citations. However, little is known about the production of data, field-specific practices, and other basic requirements such as the format a data record should have to facilitate information retrieval and bibliometric analyses. Previous studies focusing on Thomson Reuters' Data Citation Index (now Clarivate Analytics) have explored disciplinary biases and data types included (Torres-Salinas, Martín-Martín, & Fuente-Gutiérrez, 2014), data citation practices between fields (Robinson-García et al., 2015), and the relation between data citations and data mentions in social media (Peters, Kraker, Lex, Gumpenberger, & Gorraiz, 2016).

In a recent report, Costas et al. (2013) highlighted the need for developing data publication standards, reducing the dispersion of data repositories, and facilitating the traceability, citation and measurement of data records. The most comprehensive source for open data currently available is DataCite, which contains more than 7 million freely accessible records, almost doubling the figures last reported for the Data Citation Index (Peters et al., 2016).

In line with the open science movement and calls for increased data sharing and reuse, we highlight the importance of data publications and citations. This paper analyzes the structure and type of metadata offered by DataCite to assess its potential to become an important source for developing data-level metrics. DataCite is an international non-profit organization formed in 2009. It is a consortium of public research institutions, funding bodies and publishers worldwide whose mission is to promote open research data accessibility and tracking. For the latter, DataCite advocates for the use of Digital Object Identifiers (DOI) by assigning DOIs to their records (DataCite, 2015).

## 2. Objectives

This paper aims to explore the characteristics of the data collected by DataCite to determine its potential as a new source of bibliometric data for the study of open data production. Specifically, we examine the database structure and the level of standardization of the information provided in each field, to assess the usability of the data for bibliometric purposes. The paper is structured as follows. Firstly, we present the metadata scheme of DataCite records (2015). Then we assess the completeness of the data in each specific field and give an overview of the database coverage. Finally, we discuss the potential of DataCite as a source for tracking open data production, and we provide some recommendations for its use as tool for studying data production and citation patterns.

## 3. Data and methods

This section is structured in three parts. The first describes the different points of access available by DataCite and advantages and limitations of using one or the other. Second, we recollect and describe the information provided by DataCite as to its structure, definition of data record fields, and information requested to each repository. The aim is to give the reader a full account as to what DataCite expects to receive from each data repository and how this information is expected to be presented to the final user. The last part describes the dataset downloaded from DataCite's public OAI API. The information retrieved and its structure is compared with the information provided in the first subsection.

### 3.1 Points of access to DataCite

DataCite provides to APIs to the public for downloading records indexed in its database. These two points of access contain the same number of records but differ in the structure in which they are presented as well as in the detail of information provided.

**DataCite Metadata Store (https://oai.datacite.org/).** The DataCite Metadata Store is a service to manage activities related to Digital Object Identifier (DOI) registration at DataCite. The MDS is used to create, register, store and manage DOIs and associated dataset metadata created by DataCite's users and members. Here we are presented to raw data as provided by DataCite's members and has not yet been processed by DataCite.

**DataCite REST API (https://api.datacite.org/).** The DataCite REST API includes the same contents as the DataCite Metadata Store but with added layers of information by record. The DataCite team adds new information to each record regarding funding, ORCIDs, citations not provided by the data centers themselves are added.

As well as these two points of access, DataCite allows bulk queries via two additional URLs: search SOLR (https://search.datacite.org/ui) and search (https://search.datacite.org/). In this paper, we have used the DataCite Metadata Store to retrieve all records from DataCite. Throughout the rest of the paper all references made to DataCite's metadata structure are based on such information.

### 3.2 DataCite Metadata Scheme v. 3.1

In April 2016, we retrieved all records from DataCite using their public OAI API (https://oai.datacite.org). DataCite provides a metadata scheme which shows the record structure and defines each field (DataCite Metadata Working Group, 2015). Note that although a 4.0 version of the metadata scheme has recently been implemented, in this paper we refer to version 3.1 as it was the schema in place at the time of data collection. This version includes mandatory, recommended and optional fields. In the following sections, we briefly describe the main fields retrieved from the DataCite Metadata Store.

### 3.2.1 Mandatory fields

*Identifier.* While in principle DataCite encourages and promotes the use of DOI numbers, it also allows the inclusion of other unique identifiers (e.g. URN, CCDC, INCHI key, URL).

*Creator.* This field includes the name, surname or affiliation name of the creators of the data records. It would be equivalent to the author field of bibliographic records.

*Title.* The name by which the resource is known. Sometimes it also includes subtitle as a sub-field.

*Publisher.* DataCite defines publisher as "[t]he name of the entity that holds, archives, publishes prints, distributes, releases, issues, or produces the resource" (DataCite, 2015). For the current practice, there can be different interpretations on this definition thus could be performed by different actors. Hence, it can result in ambiguity on the type of entities assigned as publisher, namely individual authors, institutions, or individual data repositories. We discuss this limitation in subsection 4.2.

*Publication Year.* The year in which the data record was made publicly available, which may differ from the year of its creation. DataCite's documentation acknowledges that this can be problematic in certain cases leaving up to the user depositing the data to choose their preferred date for citation purposes.

### 3.2.2 Recommended fields

*Subject.* This is a free text field that can include keywords, classification codes, subjects, or key phrases. It includes as subfield the subject scheme used, if any, with a link to the subject scheme.

*Contributor.* This field includes the institutions and individuals involved on the collection, management, distribution or other types of contributions to the production of the data. It includes as subfield the type of contribution (i.e., contact person, data collector, etc.).

*Date.* Due to the potential ambiguity of the publication year, this field allows to specify more than one date which may be relevant for the user, such as data availability, collection, publication, etc.

*ResourceType.* Here, a two-level classification of data types is introduced. While the top level is a closed list of 15 data types, the second level classification is a free text field.

*RelatedIdentifier.* This field contains identifiers different from the DOI.

*Description.* This is a structured field. If used, free text can be entered but the type of content (abstract, methods, series information, table of contents, and other) must be specified.

*GeoLocation.* Includes the geographical location in which the data presented was collected.

## 3.3 General description of the retrieved database

Data were parsed and organized into an SQL database. A total of 7,440,415 records were retrieved. The API does not provide the recommended *Geolocation* field. This field was included in September 2016. It provides five optional fields: *Relation, Format, Language,* and *Rights.* Furthermore, the fields *Identifier* and *RelatedIdentifier* and the fields *Publication Year* and *Date* are combined in two fields (Identifier and Date). Additionally, it indicates the Data Center providing the records to DataCite. 762 organizations were included as data centers at the time of the download. These organizations have contracted with an individual DataCite member to assign DOIs. Appendix A includes a detailed description of each field retrieved and the information they contain.

Figure 1 shows the share of records in DataCite with information in each of the fields described in Appendix A. We see that many records contain empty fields (even mandatory ones). A total of 1,092,131 records (14.7% of all records collected) include no data at all. This appears to be caused by modifications made by DataCite in the data structure. More specifically, DataCite employs the Open Archives Initiative Protocol for Metadata Harvesting (OAI-PMH) and assigns an OAI id to each record. It appears that when a record needs to be modified, a new record is created with the updated information. The information in the old record is deleted (except for the OAI and the data center information), but not the record itself. Figure 2 shows an example of an empty record. This is an important element to consider when working with DataCite's API as these records should be removed from the sample.

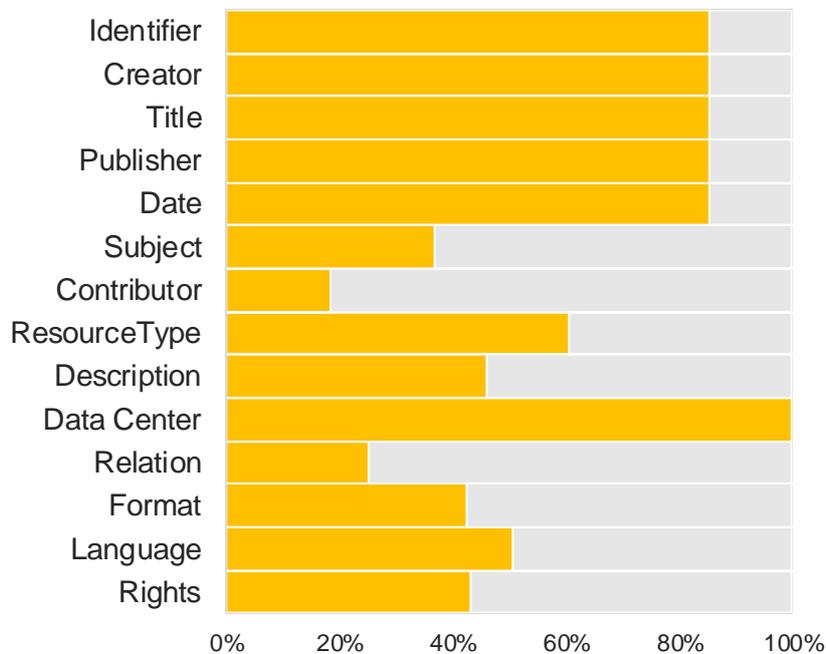

*Figure 1. Distribution of metadata information by fields*

When focusing on the records that do include information (6,348,284 records), we still find that 1,306 records (0.02%) do not include a title or publisher information. Resource type and language are reported in 60% and 51% of the records, respectively. The contributor (18%) and relation (25%) fields have the lowest presence in DataCite records.

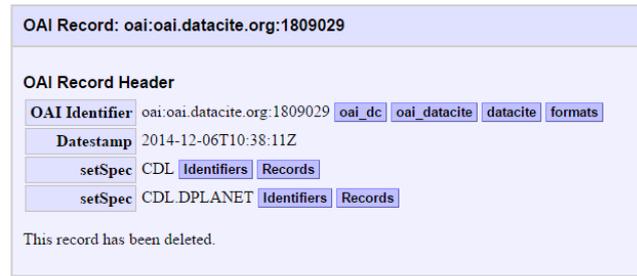

*Figure 2. Example of an empty record retrieved from DataCite's API*

# 4. Results

In this section, we report our findings regarding the content of each field and the level of standardization of the data. First, we present descriptive statistics on different types of data records. Then we analyze the geographical distribution of data centers and the number of records by country. We also analyze the publisher field to disentangle the different types of entities it contains. We also present an overview of the different types of dates included in the database. Finally, we focus on the description of the relation field, which contains DOIs of related records, trying to understand the type(s) of linkages captured by DataCite.

## 4.1 Resource types

The *ResourceType* field presents a controlled list of 15 values, complemented by a free-text subtype. Table 1 reports the total number of records by resource type and the three most common subtypes. We observe that 42% of the records are categorized as datasets, following by text (18%), image (14%), and collection (7%). As observed in table 1, most of records with a *ResourceType* 'text' are manuscripts, conference papers or journal articles. Records tagged as images are heterogeneous, ranging from academic posters to historical manuscripts, or data figures. The subtype is not mandatory and is thus empty in many records. For instance, only 4.3%, 6% and 6% of records with the resource type "Model", "Sound" and "Film", respectively, have a subtype. Overall, we find 158,781 different variations of resource subtypes, a natural off-shoot of it being a free-text field, but which reflects different understandings of what is data and what is included by each of the 15 data types.

*Table 1. Records by resource type and share of top 3 most common subtypes in DataCite. In bold-cursive subtypes appearing in more than one data type category*

| Resource type | Number of records N | % | Most frequent subtypes |
|---|---|---|---|
| Dataset | 1,867,627 | 41.69 | Dataset (63.5%), Metadata (5.8%), Data package (4.1%) |
| Text | 786,882 | 17.56 | Conference papers (15.5%), Journal articles (15.4%), ***Report*** (10.1%) |
| Image | 641,404 | 14.32 | Image (11.9%), Figure (11.2%), Plate (8.1%) |
| Collection | 303,638 | 6.78 | Collection (20.7%), Gaussian job archive (9.1%), ***Report*** (4.7%) |
| Software | 12,340 | 0.03 | Simulation tool (16.9%), Software (10.8%), Code (5.3%) |

| Resource type | Number of records N | % | Most frequent subtypes |
|---|---|---|---|
| Audiovisual | 4,470 | 0.10 | Audiovisual (43.8%), Media (23.9%), Teaching material (8.5%) |
| Film | 960 | 0.02 | Experiment (5.4%), Video (0.4%), Animation (0.1%) |
| Physical Object | 587 | 0.01 | Archival object (63.9%), HIAPER-HAIS airborne sensor (2.4%), Physical object (0.9%) |
| Event | 508 | 0.01 | Conference presentation (73.4%), Presentation (9.6%), Event (1.6%) |
| Model | 470 | 0.01 | Model (2.8%), Ontology (0.9%), Shapefiles (0.2%) |
| Interactive Resources | 287 | 0.01 | Interactive resources (12.2%), Learning object (2.1%), Sites Web (0.3%) |
| Sound | 234 | 0.01 | Recording, oral (4.3%), Sound (0.4%), Conference (0.4%) |
| Workflow | 209 | < 0.01 | Taverna 2 workflow (7.2%), Workflow (1.0%), RapidMiner workflow (0.5%) |
| Service | 18 | < 0.01 | Service (88.9%), S-map (5.6%), Data provider (5.6%) |
| Other | 871,549 | 19.45 | Data sheet (98.2%), Oceanographic cruise (0.7%), Field expedition (0.7%) |
| **Total** | **4,480,077** | **100** | |

We also observe classification redundancies between the two levels. For example, the resource type "dataset" has a subtype also called "dataset". There are also redundant subtypes between different resource types. For example, the subtype "report" appears as a subtype of both the resource types "collection" and "text". A specifically problematic case is the *Resource Type* "other", for which 98.2% of the records have a subtype labeled as "Data sheet". This suggests that these records could perhaps be considered as datasets. Taking a closer look at these records, we found that they were all derived from the same repository, Data-Planet. Actually, all records from Data-Planet are classified as "Data sheet". This variability in the distribution of records may reflect some inconsistencies in the way data centers classify records according to the scheme proposed by DataCite.

From now on, we will refer as "data records" to all those records in DataCite that have a resource type different than "text" (i.e. we consider as data-related records all records that are not articles such as manuscripts or pre-prints).

## 4.2 The geographic distribution of data infrastructures

In this section, we focus on the data providers and the countries in which they are based, to provide insights on how data infrastructures are being developed in different countries. DataCite provides a closed list of 762 institutions from which records are retrieved. The distribution of records across these data centers is uneven: 15 (2%) data centers account for more than 80% of all records. Figure 3 shows the distribution of records by resource type (excluding the records where this field is empty) for the 20 data centers who provided the most records.

The data highlights the variety of institutions providing data: from thematic data repositories (Data-Planet, PANGAEA, Digital Science), to scientific social platforms (ResearchGate) or universities (Imperial College London, ETh Zürich). Data-Planet is the largest data center in DataCite, providing 20% of all the records. As mentioned before, all records provided by Data-Planet are "data sheets". Also, some data centers (ResearchGate, E-Periodica, Universität Zürich, Zora, and ETH E-Collection) provide only "text" records.

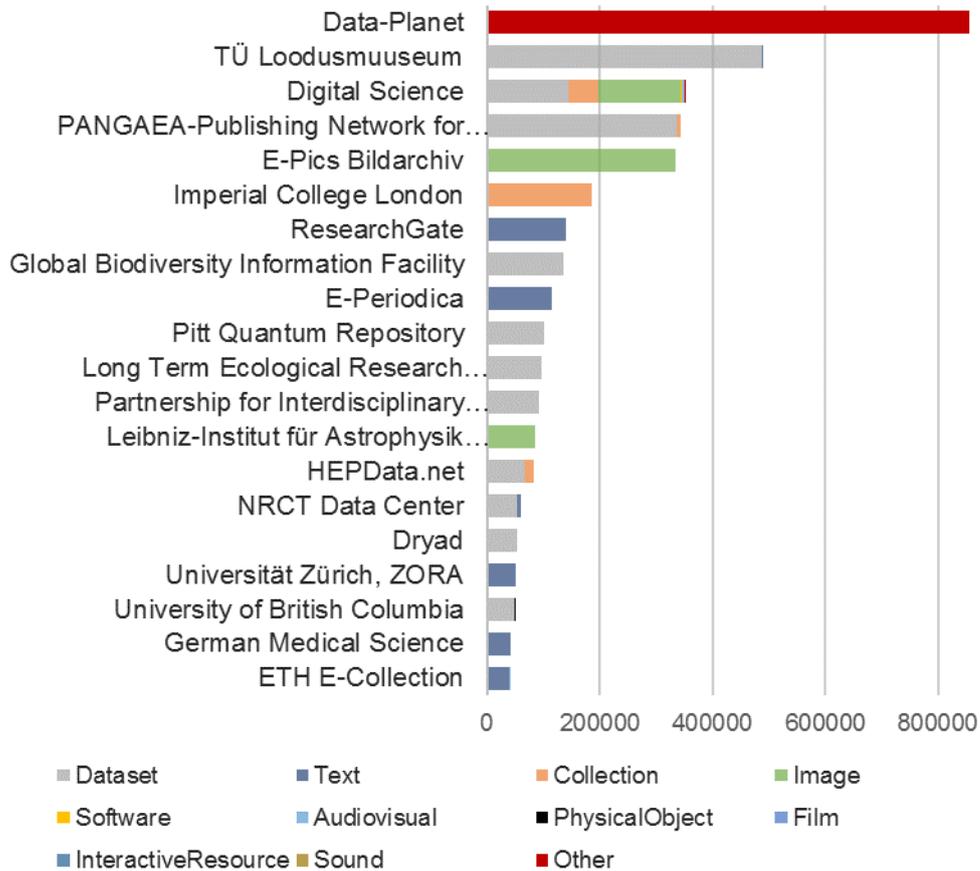

*Figure 3. Top 20 data centers by data types*

In Table 2 we assigned each data center to their countries. This information was retrieved from DataCite Statistics (https://stats.datacite.org/). It is important to note that the classification was based on the location of their headquarters, and that some data centers were associated to more than one country if they have headquarters in different countries. The country distribution in Table 2 does not reflect the affiliation of data creators nor the geographic origin of countries, but provides an overview of countries contributing towards the development of an open data infrastructure. We find that the distribution of records by country is very skewed: the United States, Germany and the United Kingdom account for 82% of the total records. The distribution of resource types also differs by country. For instance, almost 100% of records coming from Estonia, Denmark, and Canada are data records, while this proportion is much smaller in other countries such as Hungary (0.8%), Italy (4.2%), Ireland (16.1%), Australia (19.6%), and Germany (26.5%). Moreover, no data records were found in data centers based in Austria, Russia, Iran, South Korea, Liechtenstein, Slovenia, and Japan.

*Table 2. The number of data centers, number of records and share of records after excluding records labeled as data type "text" by country. Countries are ordered by total number of records*

| Countries | Data centers | # records | % data records* | Countries | Data centers | # records | % data records* |
|---|---|---|---|---|---|---|---|
| USA | 217 | 2952086 | 58.6% | Hungary | 37 | 1809 | 0.8% |
| Germany | 185 | 1795638 | 26.5% | Poland | 4 | 1713 | 1.3% |
| UK | 66 | 1382661 | 49.9% | Russia | 3 | 1388 | 0.0% |
| Switzerland | 48 | 1120868 | 32.1% | Iran | 2 | 1292 | 0.0% |
| Estonia | 6 | 489896 | 99.5% | Romania | 3 | 1032 | 47.2% |
| Denmark | 5 | 138640 | 98.0% | China | 2 | 703 | 31.2% |
| Canada | 24 | 85984 | 93.5% | Czech Republic | 1 | 470 | 100.0% |
| Thailand | 1 | 61529 | 87.9% | South Korea | 1 | 188 | 0.0% |
| Italy | 35 | 50350 | 4.2% | Belgium | 1 | 106 | 79.2% |
| Netherlands | 16 | 49900 | 80.8% | South Africa | 1 | 105 | 93.3% |
| Austria | 7 | 36450 | 0.0% | Liechtenstein | 1 | 56 | 0.0% |
| Australia | 41 | 24122 | 19.6% | Ghana | 1 | 53 | 98.1% |
| Ireland | 3 | 23181 | 16.1% | Spain | 2 | 37 | 100.0% |
| France | 32 | 13093 | 48.7% | Slovenia | 1 | 18 | 0.0% |
| New Zealand | 2 | 3081 | 39.4% | Japan | 1 | 15 | 0.0% |
| Sweden | 6 | 2835 | 97.9% | Tanzania | 1 | 10 | 90.0% |
| Unknown | 8 | 2722 | 1.8% | Uruguay | 1 | 1 | 100.0% |

* Data records are defined as all data types excluding text

The second source of information relating to open data providers is obtained from the *publisher* field. It is a non-standardized free-text field in which we found 118,136 different names. The distribution of records is highly skewed, hence by manually disambiguating the most common 1,148 publishers we managed to cover about 90% of all the records that include publisher information.

For each of these 1,148 publishers, we assigned two variables: country and type of entity. The *Country* information was retrieved from the publishers' websites and corresponds to the country where the publisher is located (like data centers, multiple countries can be assigned to a single publisher). Figure 4 presents the number of records for each country. Only records including resource type and publisher information are represented (3,704,161 records). While the distribution of records by country is similar using either the data center or publisher information, there are notable differences. We find that the number of countries contributing to DataCite is lower when using the publisher information than when using data center location. For example, no record would be assigned to Estonia, Thailand or Ireland using this method. However, they occupy the third, eighth and twelfth positions respectively when using the data center. At the other extreme, Italy, Belgium and Spain are clearly underrepresented according to data centers' location.

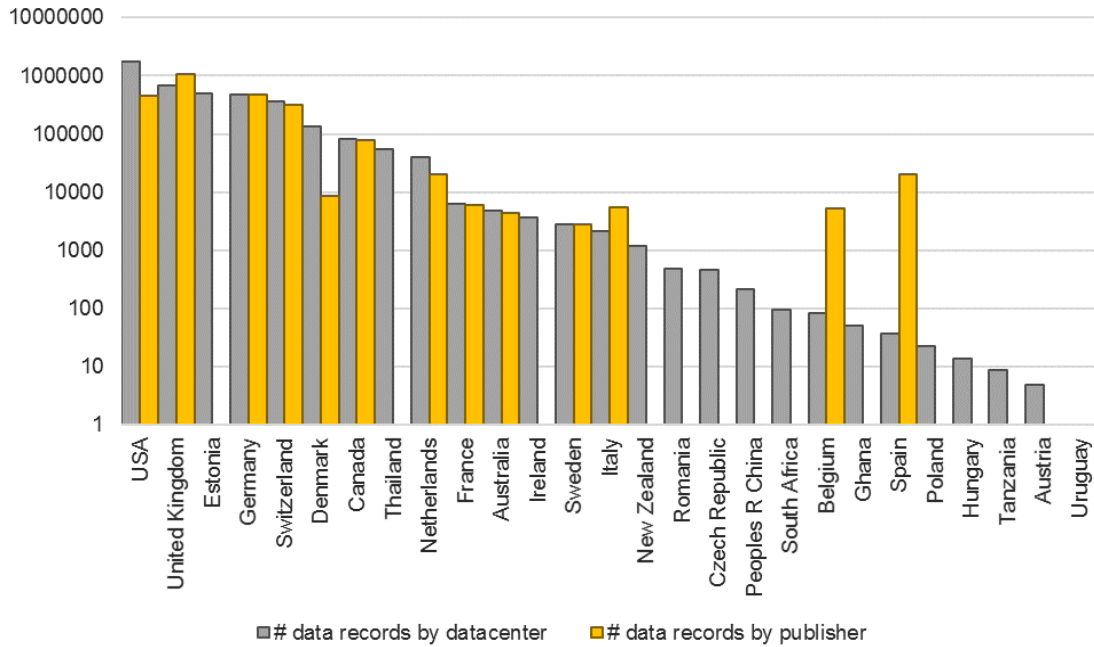

*Figure 4. Total number of data records (excluding data type "text") by country using data center and publisher affiliation data. Y-axis are logarithmic. Countries are ordered according to the total number of records using the data center affiliation.*

We also divided the publishers in 11 types of entity to better comprehend what users understand as "data publisher", but also to identify different types of institutions publish data products. We distinguish four types of repositories (i.e., national, institutional, disciplinary, and multidisciplinary repositories), and the other entities are diverse groups (research body, professional body, and educational body), publishers, firms, conferences and individuals. Appendix B provides more details on this classification.

As shown in Figure 5, a total of 156 distinct entities are identified from the 1,148 name variants disambiguated from the *publisher* field. Most of the records were assigned to 18 thematic repositories (43%). Among 156 entities, 35 are institutional repositories, followed by 33 research bodies (e.g., research centers and scientific associations), and 24 academic publishers (journals). In second and third place but with a substantially lower proportion of data records, we find institutional repositories (17%) and research bodies (15%). The proportion of data records varies substantially by publisher type. While 89% of records included in multidisciplinary repositories are data records, none of the records published by professional bodies, conferences and authors are data records. These results reflect the conceptual problem still existing on the meaning that "publishing" has in the data production model (Costas et al., 2013) or at the very least, the effect of the diversity of records included in DataCite.

| Type of publishers | # records | % data records | # publishers |
|---|---|---|---|
| Thematic repository | 2205204 | 67,78% | 18 |
| Institutional repository | 852954 | 83,26% | 35 |
| Research body | 764962 | 43,18% | 33 |
| Multidisciplinary repository | 408355 | 88,45% | 2 |
| Scientific publisher | 149305 | 9,85% | 24 |
| National repository | 40634 | 42,33% | 5 |
| Firm | 20704 | 9,67% | 6 |
| Professional body | 19215 | 0,00% | 5 |
| Conference | 18571 | 0,00% | 1 |
| Individual | 8025 | 0,00% | 2 |
| Educational body | 2326 | 93,59% | 1 |
| Not found | 621544 | 78,74% | 24 |
| **Total** | **5111799** | **72,47%** | **156** |

*Figure 5. Number of records and share of data records (after excluding text) by type of publisher. Only records with publisher information and data type are shown.*

## 4.3 Publication year and related dates

*Publication year* is a key field in any bibliometric analysis intending to provide a longitudinal perspective or to frame the study period(s). DataCite requires the *publication year* to be presented in a four-digit format. However, an important point to consider for the development of data metrics is that data records can be subjected to different actions occurring on different dates of actions, that may all be included in the metadata. Thus, DataCite (2015) has two date-related fields: *publication year* and *date*. The *publication year* field is a mandatory field that DataCite Metadata Working Group (2015) defines as "the year when the data was or will be made publicly available". Still, DataCite acknowledges that this information may be unclear or unavailable, providing alternatives such as, "[if] that date cannot be determined, use the date of registration" or "[i]f an embargo period has been in effect, use the date when the embargo period ends". Concluding that "[i]f there is no standard publication year value, use the date that would be preferred from a citation perspective".

The *date* field is an optional free-text field that can refer to different dates relevant to the record. These can be related to the date when the dataset was created, uploaded to a repository, made publicly available, updated, etc. Thus, when information if provided in the *date* field, one of the following 9 subtypes is required: accepted, available, copyrighted, collected, created, issued, submitted, updated and valid.

As mentioned before and presented in Appendix 1, the field "date" retrieved DataCite Metadata Store OAI API combines both the *publication year* and *date* in a single field. Hence the distinctions discussed above are not available. This means that multiple dates may be assigned to a single record and that the *publication year* field can only be distinguished from the *date* field when the latter is not in a four-digit format. Therefore, the date information retrieved with the API must be somehow processed before used. In this study, we define "publication year" as a date presented with a four-digit format. We identified 4,242,804 data records with this format. This cleaning process is not completely accurate as a total of 50,679 records reported publication years above 2099 or from early 1000s and were thus not considered[1]. Figure 6 shows the number of records for the 1950-2020 period. We observe many records dating from 2016 onwards due to the embargo they are restricted by.

---
[1] Although there are cases of data records dating from the early 1000s, e.g., digitalized archival objects.

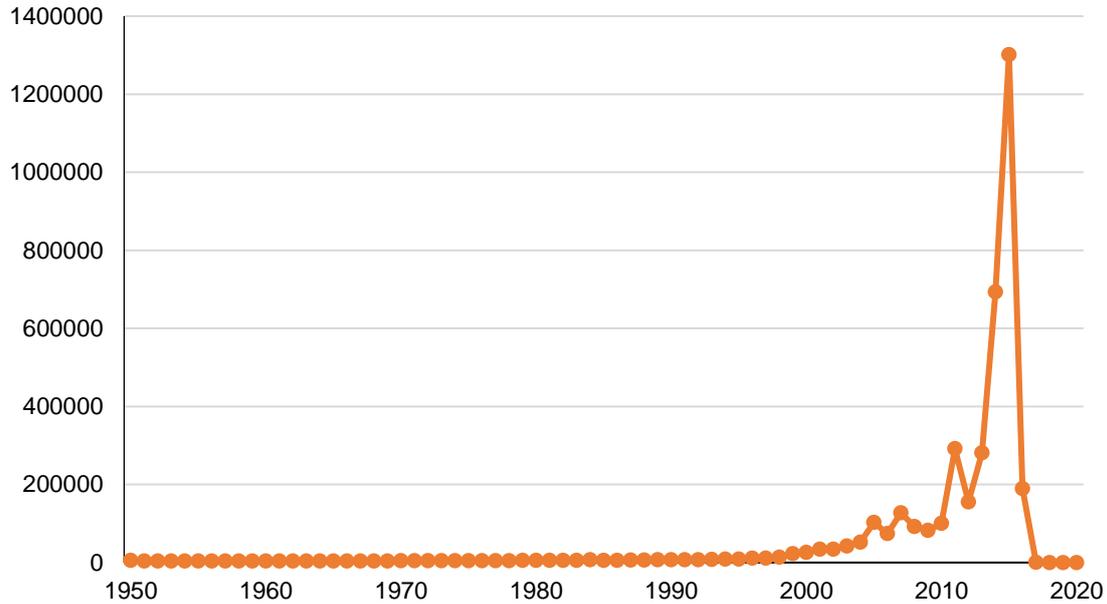

*Figure 6. Number of records per year using the publication year in DataCite. 1950-2020 period*

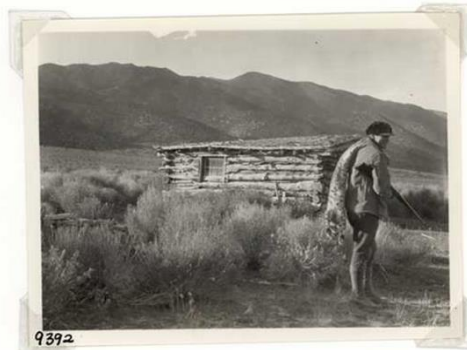

*Figure 7. Example of record with an older date to the development of data repositories. 6A. Contents of a photograph taken in 1929. 6B Data record in DataCite. The date of publication of the record is 1929.*

The fact that there is no clear definition for the *publication year* field, may lead to some discrepancies in the data. This is especially meaningful in the case of historical data where the user could choose to indicate the date of the historic record or the date of its retrieval. Figure 7 provides the example of a digitized photograph which had already been published in its physical form. Here, the *publication year* field contains the value 1929, which is in fact the date when the photograph was taken.

Regarding records including additional dates, we identified 2,095,183 records of which 43% reported the availability date, 25% reported the date of creation 14% declared the collection date and 12% an update and 3% and issue date. Less than 0.2% of the records reported the date of copyright, submission, validity or acceptance.

## 4.4 Related DOI numbers

The OAI DataCite API also provides a field named *relation*, which is equivalent to the *RelatedIdentifier* field in the DataCite Metadata Schema. The main difference is that here we retrieve only the information provided by the data centers, while the *RelatedIdentifier* field retrieved from the REST API includes additional relating provided by the DataCite team. It contains identifiers for publications (e.g., DOIs, arxiv, bibcode, handles; not necessarily in DataCite). As all records in DataCite include a DOI number along with other associated identifiers, we crossed related DOI numbers with: 1) the DataCite database itself, to find potential relations among data records within DataCite; and 2) with the Web of Science, to identify potential relations with scientific publications. As shown in Figure 8A, 23% of all DataCite records include related DOIs. The number of related DOI numbers by record varies greatly, showing a highly-skewed distribution (Figure 8B). Figure 8C crosses DataCite related DOIs with DataCite records, with DataCite records defined as datasets, and with Web of Science records. Less than 25% of the related DOI numbers belong to other DataCite records. Approximately 15% belonging to articles indexed in the Web of Science (Figure 8C). When we focus on the data type of related DOIs contained in DataCite (Figure 8D), we observe that 90% of these are datasets. After a cursory check of some of these cases, we observe that occasionally the relation is formed by a container data record (i.e., a database) and its tables (i.e., datasets). For example, the database doi:10.15468/dl.qnbifh included at the time of the data collection, 5,192 related datasets. This partially explains the skewed distribution observed in Figure 8B. In other cases, the relation indicates data (re)use by linking the data with a paper. However, this field does not seem to contain the DOI of articles citing the data record, and we find no evident criteria for characterizing the types of relations reported in this field.

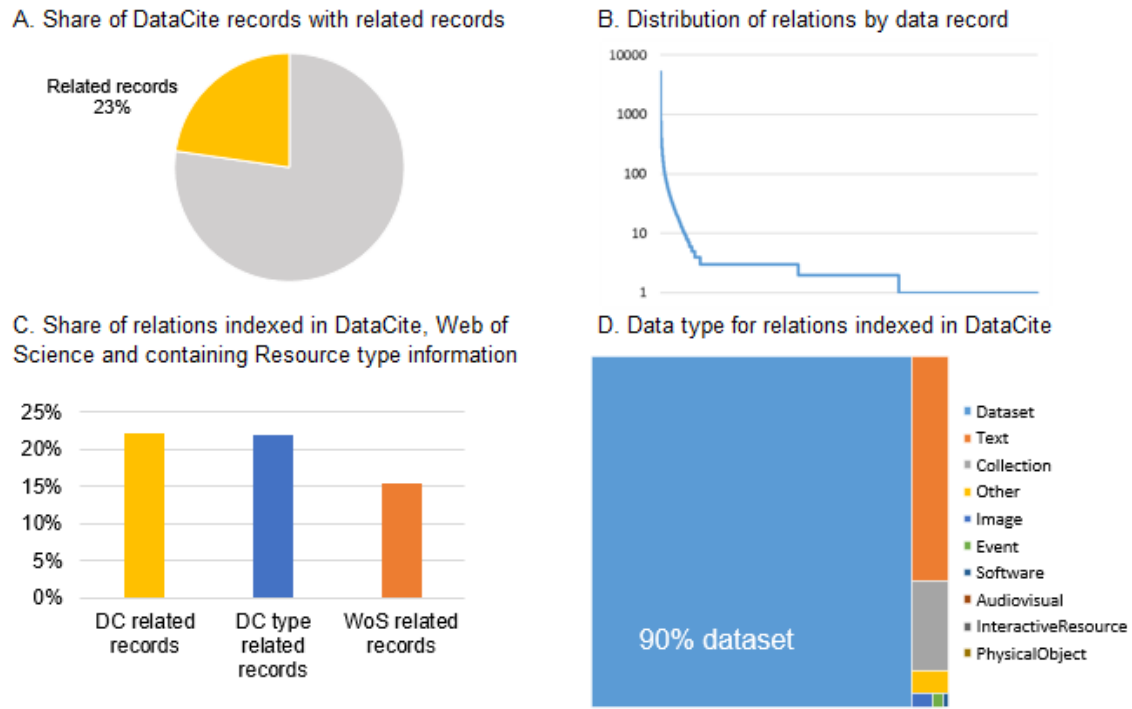

*Figure 8. Analysis of the relation field in DataCite. A Share of records in DataCite with related DOI numbers within DataCite records. B. Distribution of the number of related DOI numbers by data record. C. Share of related DOI numbers included in DataCite by their data type. D. Share of related DOI numbers indexed in DataCite, indexed in DataCite and with data type information, and indexed in Web of Science.*

Interestingly, Robinson-García et al. (2016) reported a similar type of relations also consigned in the Thomson Reuters' Data Citation Index, although in that case, only relations between datasets and scientific papers were included. However, they reported a repository dependence of the reporting of these relations, that is, depending on the repository we would find records with relations or not. In DataCite there is evidence suggesting that such a dependency also exists, in this case with data centers: only 226 (30%) data centers reported at least one data record with a related DOI number, and 44 (5%) of them reported related DOI numbers in all their records (see Figure 9).

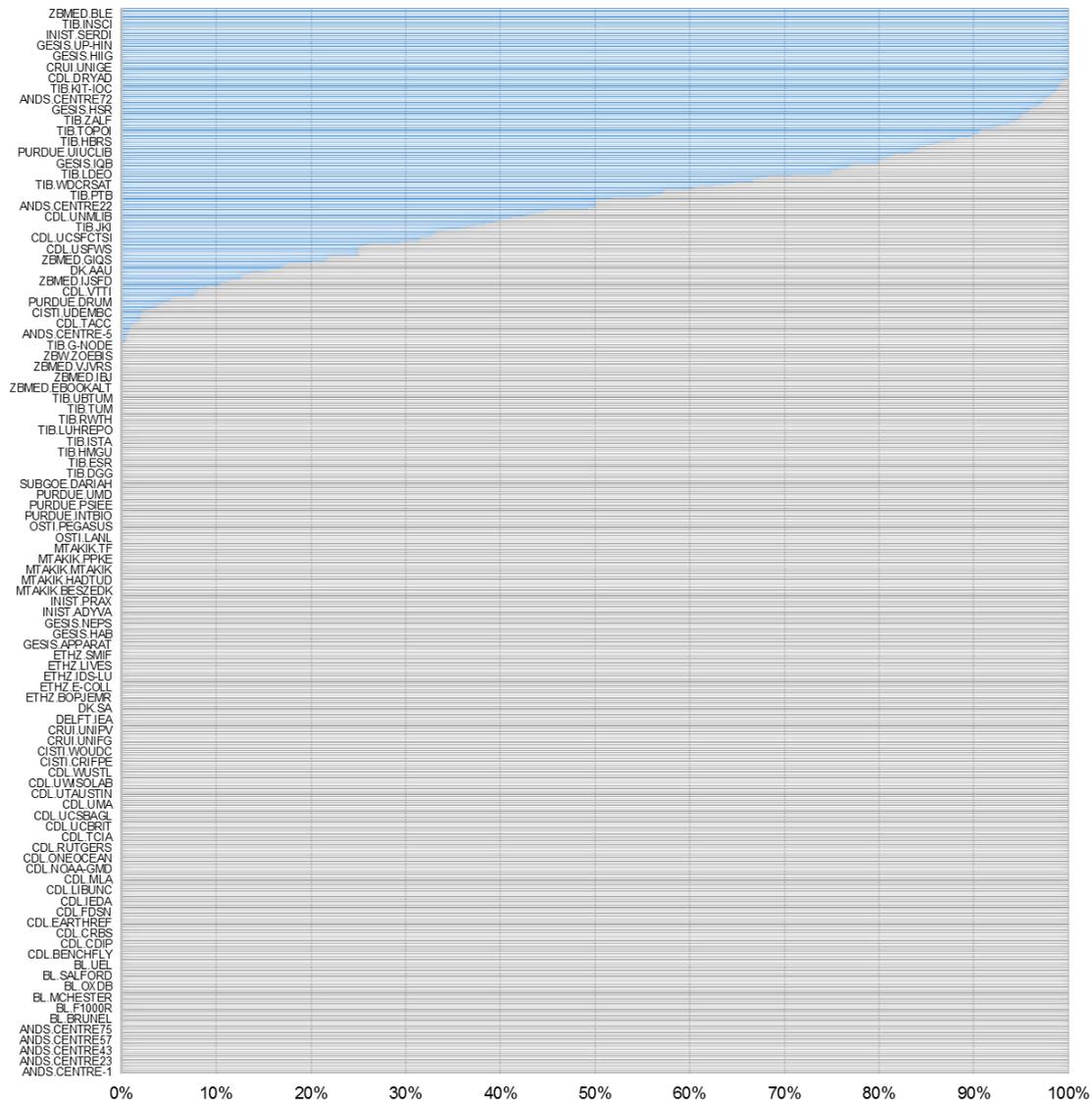

*Figure 9. Share of records with related DOI numbers assigned to them. Blue represents records with related DOI numbers. Grey represents records with no related DOI numbers reported.*

# 5. Concluding remarks and recommendations

The research on data sharing and open data is growing, while at the same time funding bodies are encouraging greater research transparency. Terms like data-driven science, data-intensive science, and open science are becoming more and more common in policy documents and statements such as the European Unions' Horizon2020 (European Commission, 2016). In this context, DataCite is called to play an important role as source for the analysis and study of data publication and reuse. While the demand of data metrics has been a constant since the beginning of the 2010s (Costas et al., 2013), there is still a long way to go until the movement expands to broader fields of Science and to more countries.

This paper presents the first large-scale data collection and analysis of DataCite to assess its potential as a bibliometric tool able to provide information and metrics about open data activities at a macro-scale. Compared with other similar products such as the Data Citation

Index, the size and richness of DataCite data offer greater possibilities as a bibliometric source for developing open data metrics. Still, this richness of data comes at a price. Conceptual problems such as what is data or to which scientific field or discipline different datasets belong to, along with technical problems such as the lack of standardization of many of its fields, may still represent an advantage towards the Data Citation Index, in which the structure of fields in the Data Citation Index adapts to some extent the structure of bibliographic records. This is presents a positive advantage for the Data Citation Index because it allows bibliometric analyses without prior processing (e.g., Robinson-Garcia, Jiménez-Contreras & Torres-Salinas, 2016). However, this analytical simplicity of the Data Citation Index overlooks some of the key issues found when exploring the nature and heterogeneity of open data. As shown in this paper, the metadata of DataCite records is very rich and heterogeneous, here we describe some of the important issues that need to be considered when using DataCite as a source of data for open data analytics.

## 5.1 Central issues regarding the metadata provided by DataCite

### 5.1.1 Data types and the definition of "data"

An important critical element that needs to be considered when working with DataCite is that as such, all records included in the database are not strictly data-related. For example, more than 12% of the valid records in DataCite are text or articles. Therefore, in order to properly identify and analyze the production of data, diverse filters need to be applied by types of data. However, we have highlighted the important diversity of data types included in DataCite. In a way, the many types of data covered in DataCite suggest that a broader understanding of what constitutes research data is very necessary. In fact, the presence of multiple data related types such as "Images", "collection" or "software" reinforces the idea that we need to stop considering "data" as a homogeneous publication type.

### 5.1.2 DataCite Metadata fields

The DataCite schema closely aligned with Dublin Core, which allows interoperability between different platforms and record types as well as ensuring minimum levels of quality of author-generated metadata (Greenberg et al., 2002). However, the simplicity of the model (Lagoze, 2001) leaves room to ambiguity in many of the fields required in order to develop any type of bibliometric analysis. We found that a major issue existing on DataCite is that a lot of records are missing information in many of the fields (even mandatory ones). In addition, making some of the recommended fields mandatory (e.g., the subject, the institutional affiliation of the creator) would enhance DataCite's potential for bibliometric analyses. It would also be useful to make mandatory a "type of relation" subfield for the "Relation" field which is one of the most promising fields for the development of data metrics. It is worth noting that this information is now available in the new metadata scheme and through the DataCite search webpage.

Moreover, the problems raised when analyzing the information provided by the *publication year* and *date* fields raises questions as to when data are produced and disseminated. Regarding the "Publisher" field, it seems that its current definition is too broad, as there is variety of entities that can hold, own, archive, publish (and so on) a digital object available in DataCite. At it has been shown, the field combines a huge diversity of entities that are not strictly publisher (e.g., repositories, research bodies, firms, etc.). In fact, since the "Data Center" information is unique for each data record, it could make more sense to use it for citation purposes rather than the publisher, which is a free text field[2].

# 6. Recommendations

Based on the results of this paper it is possible to suggest a series of recommendations that might be useful for users who wish to employ DataCite for developing data metrics, and for DataCite as a provider of data records on data sharing activities. These recommendations are intended to maximize their efforts to provide a service that efficiently promotes data publishing and data citation. The size of DataCite and the fact that it is accessible for free highlight its potential to become a valuable source of information for quantitative analyses of data production, sharing and (re)use. However, there are critical issues regarding the structure and cleanliness of DataCite records that would need to be addressed to improve its usability. In any case, the conclusions drawn here are based on the DataCite Metadata Store and do not consider any improved functionalities available through the DataCite REST API. In this sense, the advantages and limitations of using different points of access should be made clearer so that users can choose one or the other depending on the analysis they wish to conduct.

In this sense, potential users of DataCite should consider the following issues: First, empty records should be removed before attempting to make any statement regarding the actual data contained by DataCite. As noted in subsection 'General description of the retrieved database', over 1 million records were found empty at the time of the retrieval of the data. The non-removal of these records may mislead the counts of the actual size of the database.

Second, issues related to data completeness reduce the analyzable dataset as more filters are used to retrieve records. For example, to focus on only data-related records (e.g. datasets) it is necessary to filter by *ResourceType*. However, this field is empty for a substantial amount (40%) of records. In addition, the DataCite Metadata Store contains a wide variety of "resource types". Thus, users must decide before hand which data types are relevant for the analysis and understand the potential losses of information that the filters will impose.

Third, a considerable amount of data processing and cleaning will most likely be needed, as most fields are not standardized. Furthermore, the fact that some fields are merged (e.g. *publication date* and *date*) makes it compulsory to process and clean the data before analyzing it.

Finally, an important issue critical for the potential usability of the database for metric purposes is the lack of standardization of many metadata fields. Having many free text fields (e.g. *Publication year*, *publisher*, *creator*) makes data retrieval more arduous and makes it necessary to disambiguate the data. By simply imposing a standard format for certain fields such as the *creator* field, or by including a closed list for the *ResourceType* field and subfield or for the *subject* field would greatly improve the quality of the data and facilitate its analysis.

## 6.1 Further research

DataCite is currently one of the main data sources available for the development of data metrics, and a great promoter of data sharing and reuse. Indeed, despite its recent creation, DataCite is probably the largest database, with a vast and heterogeneous set of data records, bringing us a step closer to an ideal of open science characterized by its transparency and its capacity to optimize the use of resources. By providing an overview of the structure and content of the DataCite records, this paper has hopefully served as a first step towards a better understanding of data production, publication and reuse by the scientific community. Further research will focus on comparisons with different of access to DataCite records, the study of the relationships between authors of scientific publications and creators of datasets, the development of suitable classifications of data records and the presence of mentions to DOIs in the references of scientific publications to data.

---

[2] The current recommended data citation format from DataCite is the following. Creator (Publication year). Title. Publisher. Identifier (DataCite, 2015).

# Acknowledgements

Preliminary results of this paper were reported at the 3:AM Conference held in Bucharest (Romania), 27-29 September, 2016. The authors would like to thank Henri de Winter from CWTS for helping in the retrieval of the data and Kristian Garza from DataCite for fruitful and helpful discussions on points of access to DataCite and structure of records. The two anonymous reviewers are also thanked for their constructive comments and recommendations. This study has been partially supported by the European Commission project RTD-B6-00964-2013 *Monitoring the evolution and benefits of Responsible Research and Innovation* (MoRRI). Nicolas Robinson-Garcia is currently supported by a Juan de la Cierva-Formación grant from the Spanish Ministry of Economy and Competitiveness.

## *Appendix A. Retrieved fields and description of their contents*

| Field | Description |
|---:|---|
| **Identifier** | Unique number identifier. DataCite assigns DOIs to all data records, although many include additional identifiers such as CCDC (Cambridge Crystallographic Data Centre) or InChI (International Chemical Identifier). |
| **Creator** | Author of the data record. This field is not presented in a standardized format (i.e. Surname, Initials). |
| **Title** | Name of the data set or file stored in the repository. |
| **Publisher** | Non-standardized format which includes a great variety of different entities raging from repositories, journals, institutions, etc. |
| **Date** | This field includes the mandatory field 'Publication Year' as well as the 'Date' field, which means that each record can have more than one publication year. The format is standardized but heterogeneous. Hence 'Publication Year' information appears as a four-digit number while Date appears stating the type of date and the actual year (i.e., Available:01/2/2005). |
| **Subject** | Keywords assigned to each data record. While we observe that for some repositories a fixed classification system is employed, this is not systematized for all data records. |
| **Contributor** | Individuals and institutions collaborating on the creation of the data but not considered as creators. As with the 'Creator' field, this field is not presented in a standardized format. |
| **ResourceType** | This field includes both, the first-level data type classification as well as the second-level data type classification. |
| **Description** | This field includes in its content the five distinct subsections described by DataCite. However not all records include all subsections. |
| **Data Center** | Institution in charge of feeding DataCite with records. Data centers have a unique identifier each constructed in two parts. First the intermediary institution and secondly, the sending institution. For instance, BL.IMPERIAL is the identifier for Imperial College London. BL stands for British Library, the intermediary institution and IMPERIAL for the sending institution. |
| **Relation** | This field related each data record with additional DOI numbers. How such relation is established is not formally declared in the record. Despite DataCite offers a controlled list of values indicating the type of relation established between records, we did not find this information in the data retrieved. More on this in subsection 3.4 |
| **Format** | Non-standardized field which includes a formal description of the contents of the record. Here we find information which ranges from a catalographic description of the contents (i.e., Zwei Teile in 1 Band ; 17 cm) to actual format of the submitted file (i.e., SPSS file). |
| **Language** | Non-standardized field indicating the language of the record. Language is indicated by using a two-digit format, a three-digit format or the full name. In some cases, more than one language is reported (i.e., fr-en) |
| **Rights** | Non-standardized format including the holder of the copyrights if any or the license by which the data record is protected. Information is reported here not only in English but also in other languages. |

## *Appendix B. Classification of publisher types*

Publishers were classified into eleven mutually exclusive categories to analyze different national data infrastructures. Following we include the twelve types of publishers identified along with examples for each of them.

| Publisher type | Examples | # records |
|---:|---|---:|
| Thematic repository | Data-Planet™ Statistical Ready Reference by Conquest Systems, Inc.; Cambridge Crystallographic Data Centre | 2,205,204 |
| Institutional repository | Imperial College London, ETH-Bibliothek Zürich, Bildarchiv, University of Pittsburgh | 852,954 |
| Research body | Partnership for Interdisciplinary Studies of Coastal Oceans (PISCO), Leibniz Institut für Astrophysik Potsdam (AIP) | 764,962 |
| Multidisciplinary repository | Figshare, ZENODO | 408,355 |
| Scientific publisher | German Medical Science GMS Publishing House, Zofinger Tagblatt, PeerJ | 149,305 |
| National repository | Digital Repository of Ireland, Colchester, Essex: UK Data Archive | 40,634 |
| Firm | Huber & Co. AG, Verlegergemeinschaft Werk, Bauen + Wohnen Bauen + Wohnen GmbH | 20,704 |
| Professional body | Bund Schweizer Architekten, Freidenker-Vereinigung der Schweiz, Union syndicale Suisse | 19,215 |
| Conference | European Congress of Radiology | 18,571 |
| Individual | W. Jegher & A. Ostertag, J.F. Boscovits | 8,025 |
| Educational body | nanoHUB | 2,326 |